\documentclass[twocolumn,showpacs,pre,english,showkeys,10pt,aps,unsortedaddress,superscriptaddress]{revtex4-1}
\usepackage{graphicx,amsmath,amssymb}
\usepackage[english]{babel}
\usepackage{hyperref}

\begin{document}

\title{Universal critical behavior of the $2d$ Ising spin glass}

\author{L.~A.~Fernandez}
\affiliation{Depto. de F\'{\i}sica Te\'orica I. Facultad de Ciencias
  F\'{\i}sicas. Universidad Complutense de Madrid. 28040 Madrid. Spain.}

\affiliation{Instituto de Biocomputaci\'on y
  F\'{\i}sica de Sistemas Complejos (BIFI), 50018 Zaragoza, Spain.}

\author{E.~Marinari} \affiliation{Dip. di Fisica and INFN--Sezione di
  Roma 1, Universit\`a La Sapienza, P.le A. Moro 2, I-00185 Rome, Italy.}
\affiliation{Nanotec-CNR, UOS Roma, Universit\`a La Sapienza, P. le
  A. Moro 2, I-00185, Rome, Italy.}

\author{V.~Martin-Mayor}
\affiliation{Depto. de F\'{\i}sica Te\'orica I. Facultad de Ciencias
  F\'{\i}sicas. Universidad Complutense de Madrid.  28040 Madrid. Spain.}
\affiliation{Instituto de Biocomputaci\'on y
  F\'{\i}sica de Sistemas Complejos (BIFI), 50018 Zaragoza, Spain.}

\author{G.~Parisi} \affiliation{Dip. di Fisica and INFN--Sezione di
  Roma 1, Universit\`a La Sapienza, P.le A. Moro 2, I-00185 Rome, Italy.}
\affiliation{Nanotec-CNR, UOS Roma, Universit\`a La Sapienza, P. le
  A. Moro 2, I-00185, Rome, Italy.}

\author{J.~J.~Ruiz-Lorenzo}
\affiliation{Depto. de F\'{\i}sica and Instituto de Computaci\'on
  Cient\'{\i}fica Avanzada (ICCAEx), Univ. de Extremadura, 06071
  Badajoz, Spain.}  \affiliation{Instituto de Biocomputaci\'on y F\'{\i}sica
  de Sistemas Complejos (BIFI), 50018 Zaragoza, Spain.}

\date{\today}

\begin{abstract}
We use finite size scaling to study Ising spin glasses in
two spatial dimensions. The issue of universality is addressed by
comparing discrete and continuous probability distributions for the
quenched random couplings. The sophisticated temperature dependency of
the scaling fields is identified as the major obstacle that has
impeded a complete analysis. Once temperature is relinquished in favor
of the correlation length as the basic variable, we obtain a reliable
estimation of the anomalous dimension and of the thermal critical
exponent. Universality among binary and Gaussian couplings is
confirmed to a high numerical accuracy.
\end{abstract}


\maketitle

\section{Introduction.}
Spin glasses \cite{edwards:75} are a rich 
problem~\cite{binder:86,mezard:87,fisher:91,young:98,mezard:09,binder:11b}.
In particular the Ising spin glass in $D=2$ spatial dimensions poses
questions of interest both for theory and for experiments. The system
remains paramagnetic for any temperature $T>0$, but the critical limit
at $T=0$ has puzzled theorists for many years
\cite{kirkpatrick:77,morgenstern:80,blackman:82,mcmillan:83,cheung:83,bhatt:87,singh:86,wang:88,freund:88,freund:89,blackman:91,berg:92b,saul:93,saul:94,rieger:96,rieger:97,hartmann:01b,carter:02,amoruso:03,lukic:04,jorg:06b,lukic:06,liers:07,katzgraber:07b,parisen:10,thomas:11,parisen:11,jorg:12,lundow:15a}.
On the other hand recent experiments in spin glasses are carried out
in samples with a film geometry
\cite{guchhait:14,guchhait:15a,guchhait:15b}. The analysis of these
experiments will demand a strong theoretical command.

In the limit $T\to 0$ the physics of the system is dictated by the
low energy configurations of the system. The nature of the coupling
constants $J$ becomes the ruling factor: if the $J$ are discrete and
non vanishing, an energy gap appears. Instead, the gap disappears if the
couplings  are allowed to approach with continuity the value $J=0$. Several
Renormalization Group (RG) fixed points appear at $T=0$, depending on the
nature of the
couplings distribution~\cite{amoruso:03}. However, most of these fixed
points are unstable even for the tiniest positive temperature: the
only remaining universality class is the one of the continuous
coupling
constants~\cite{jorg:06b,parisen:10,thomas:11,parisen:11,jorg:12} (the
very same effect is found in the Random Field Ising
model~\cite{fytas:13}).

The distinction between universality classes is unambiguous only in
the thermodynamic limit. For finite systems of size $L$, samples with
discrete couplings display a crossover at scale $T^*_L$ between
continuous ($T\gg T^*_L$) and discrete behavior ($T\ll T^*_L$).  How
$T^*_L$ tends to zero for large $L$ has been clarified only
recently~\cite{thomas:11,parisen:11} (see below).

Perhaps unsurprisingly given these complications, the critical
exponents of the model are poorly known. For the thermal exponent
$\nu$ ($\xi \propto T^{-\nu}$, where $\xi$ is the correlation length)
we only have crude estimates, $\nu\approx 3.5$~\cite{jorg:06b}
(estimates can be given by using indirect methods, see
below). Even worse, the anomalous dimension $\eta$ has been till date
impossible to estimate \cite{jorg:06b,katzgraber:07b,parisen:11}
(correlations decay with distance $r$ as $C(r)\sim 1/r^{D-2+\eta}$ for
$r\lesssim \xi$, making $\eta$ crucial for an out of equilibrium
analysis~\cite{janus:08b,janus:09b,fernandez:15}). Besides, little is
known about corrections to the scaling exponent $\omega$.

Here, we remedy these state of affairs by means of large scale Monte
Carlo simulations. Crucial ingredients are: (i) we consider both
continuous and discrete coupling distributions; (ii) multi-spin coding
methods (novel for Gaussian couplings) provide very high statistics;
(iii) the non-linear scaling fields (whose importance was emphasized
in Ref.~\cite{hasenbusch:08}) cause severe problems in the finite size
scaling close to $T=0$, that we are able to solve~\footnote{When the
  critical temperature, $T_\mathrm{c}$, is nonzero the problems caused
  by the non-linear scaling fields can be bypassed using a standard
  analysis~\cite{amit:05,nightingale:76,ballesteros:96}. In fact in
  $3D$ spin glasses~\cite{janus:13} one compares data from different
  system sizes at the \emph{same} temperature, namely $T_\mathrm{c}$,
  which cures most of the problems.}.  We also obtain for the first
time a precise numerical bound for the anomalous dimension,
$|\eta|<0.02$. This strongly supports the conjecture $\eta=0$.
Decisive evidence for universality follows from our computation of
$\omega$. For Gaussian couplings we also obtain a precise
estimate of $\nu$.

\section{Model and observable quantities.}
We consider the Edwards Anderson model on a square lattice of linear
size $L$, with periodic boundary conditions, nearest neighbors
interactions and Ising spins $\sigma_{\boldsymbol x}=\pm 1$.  The
coupling constants $J_{\boldsymbol x\boldsymbol y}$ are quenched
random variables. A \textit{sample} is a given couplings
realization. Thermal averages for a given sample are denoted as
$\langle\ldots\rangle$. The statistical average of thermal mean values
over the couplings is denoted by an over-line. We consider two
different kinds of coupling distributions, $J_{\boldsymbol
  x\boldsymbol y}=\pm 1$ with $50\%$ probability, and a Gaussian
distribution with zero mean and unit variance.  For later use, we note
a \emph{temperature symmetry}: in our problem $T$ and $-T$ are
equivalent because of the symmetry $J\leftrightarrow -J$ of the
couplings distribution.

We consider real replicas: couples of spin configurations
$\{s_{\boldsymbol x}\}$ and
$\{\tau_{\boldsymbol x}\}$ evolving with the same  couplings, but
otherwise statistically independent. Let $q_{\boldsymbol x}=s_{\boldsymbol
  x}\tau_{\boldsymbol x}$. The order parameter $q$ and the Binder ratio
$U_4$ are
\begin{equation}\label{eq:Binder}
 \textstyle q=\sum_{\boldsymbol x}q_{\boldsymbol x}/L^2\,,\qquad U_4 =
  \overline{\langle q^4 \rangle}/\overline{\langle q^2 \rangle}^2\,.
\end{equation}
$G(\boldsymbol r)= \sum_{\boldsymbol{x}} \overline{\langle
  q_{\boldsymbol{x}} q_{\boldsymbol{x}+\boldsymbol    r}\rangle}/L^2$
is the overlap-overlap correlation function.
From its Fourier transform $\hat G(\boldsymbol k)$ we
compute the spin glass susceptibility $\hat G(\boldsymbol k =0)= L^2
\overline{\langle q^2\rangle}$ and the second moment correlation length
$\xi_L$~\cite{cooper:82,palassini:99,ballesteros:00,amit:05}.

\section{Finite Size Scaling.}

Exactly at $T=0$ our two models behave very differently.  In the
Gaussian case, barring zero measure exceptions, the ground state (GS)
is unique with a continuous spectrum of excitations. As a consequence,
at $T=0$ and for any size $L$, $\overline{\langle q^2\rangle}=1$. It
follows that the anomalous dimension exponent $\eta=0$ and, according
to our definition, $\xi_L=\infty$, even for finite $L$.

The $J\!=\!\pm 1$ model is gapped, with a highly degenerate GS.  At large
distances the correlation function behaves as $G(\boldsymbol r,
T=0)\sim q_{\mathrm{EA}}^2 + A/r^{\theta_S}$, implying $\xi_L\sim
L^{\theta_S/2}$.  $\theta_S\approx
1/2$~\cite{thomas:11,saul:93,lukic:06} is the entropy exponent.  This
$T=0$ behavior extends up to the crossover scale $T^*_L\sim
L^{-\theta_S}$~\cite{thomas:11}. In fact,
Eqs.~(\ref{eq:FSS-xi},\ref{eq:FSS}) below apply for this model only
down to $T\sim L^{-1/\nu}\gg L^{-\theta_S}$~\cite{parisen:11}.

The singular part of the disorder averaged free energy
scales as
\begin{equation}
  F_{\mbox{singular}}\left(\beta,h,L\right) \simeq
  L^{-D} f\left( u_h L^{y_h}, u_T L^{y_T}  \right)\;,
\end{equation}
plus sub-leading terms. Here $u_h$ and $u_T$ are the scaling
fields~\cite{salas:00,amit:05,hasenbusch:08} associated respectively
with the magnetic field $h$ and with the temperature $T$ (since our
$D=2$ system is only critical at $T=0$)\footnote{The relationship
  between $h$ and the ``magnetic field'' $h_q$ coupled to the spin
  overlap is $h_q = h^2+{\cal O}(h^4)$.}.  The scaling fields $u_T$
and $u_h$ are (asymptotically $L$-independent) analytic functions of
$h$ and $T$ that will enter our analysis through the numerical
determination of observables like $\xi_L/L$, $U_4$, $q^2$, \dots
Recalling the $T\leftrightarrow -T$ symmetry, one can expand by
obtaining $u_T(T,h)=\hat u_T(T)+{\cal O}(h^4)$, where $\hat
u_T(T)\simeq u_1 T(1+\ u_3 T^2 + {\cal O}(T^4))$, and $u_h(T,h)=h^2
\hat u_h(T)+{\cal O}(h^4)$ with $\hat u_h(T)=c_0+\ c_2 T^2 + {\cal
  O}(T^4)$.

In terms of the scaling fields the correlation length behaves as
\begin{equation}\label{eq:FSS-xi}
\xi_L=L\, F_\xi (L^{1/\nu}\hat u_T)\ +\ {\cal O}(L^{-\omega})\;,
\end{equation}
where at variance with $\hat u_T$ and $\hat u_h$, the critical
exponents $\nu$ and $\omega$ and the scaling function $F_\xi$ are
universal~\footnote{The universality of the scaling functions in $D=3$
  spatial dimensions was carefully analyzed in~\cite{jorg:06}.}.  We
follow Refs.~\cite{parisi:80d,caracciolo:95,caracciolo:95b} and we
factor out the temperature dependency, finding:
\begin{equation}\label{eq:FSS}
  \overline{\langle q^2\rangle} = [\hat u_h(T)]^2
  F_{q^2}(\xi_L/L)\,,\ U_4= F_{U_4}(\xi_L/L)\,.
\end{equation} 
In Eq.~\eqref{eq:FSS} we have neglected again corrections
of order $L^{-\omega}$. The scaling functions $F_{q^2}$ and $F_{U_4}$
are universal.

\section{Simulation details.}
High statistics was collected using 128-bits multi-spin coding (see
\cite{newman:99} and appendix~\ref{sect:MSC-GAUSS}). In the Gaussian
case, the same bonds in the 128 copies of the system share the same
absolute value of the couplings (only sign are at random and
independent in different samples). Still, as shown in
appendix~\ref{sect:effective-number}, the statistical gain is
significant. We have equilibrated~\footnote{The elementary Monte Carlo
  step consisted of 10 Metropolis sweeps at fixed temperature,
  followed by a cluster update~\cite{houdayer:01} and by a parallel
  tempering step \cite{hukushima:96,marinari:98b}.  We consider two
  sets of two real replicas for each temperatures. The cluster updates
  are performed only within each set (overlaps are computed by taking
  a pair of statistically independent configurations, each from one
  set).  We performed a stringent equilibration test, that takes into
  account the statistical correlation when comparing the last
  logarithmic bins~\cite{fernandez:08b}.} lattices of linear size
$L=4,6,8,12,16,24,32,48,64,96$ and $128$ (see Figure 1 and
appendix~\ref{sect:parameters}).

\section{On Universality.}
\begin{figure}[t!]
\centering \includegraphics[height=\columnwidth, angle=270]{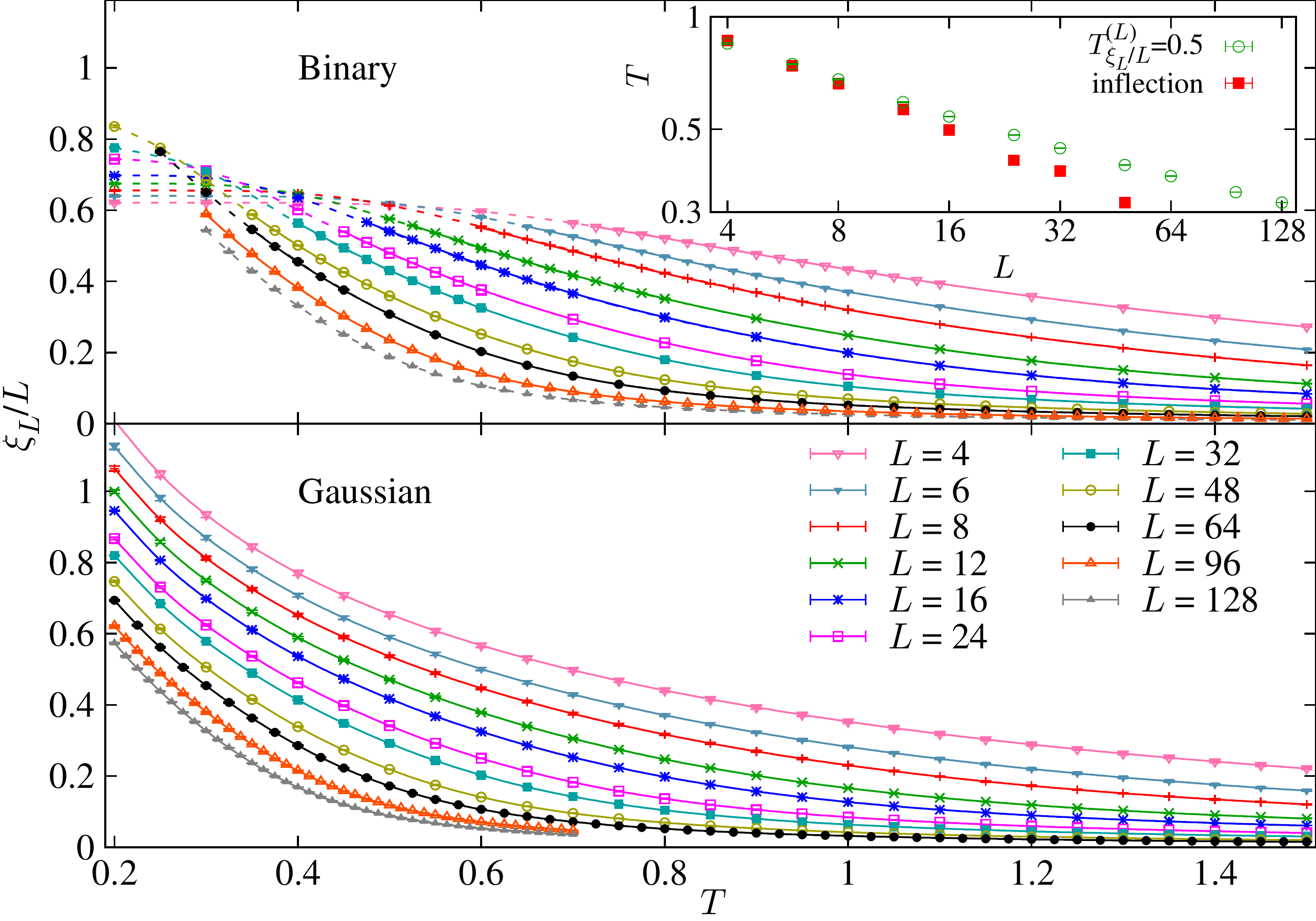}
\caption{(color online) {\bf Top:} Binary model correlation length (in
  units of the system size) versus temperature. $\xi_L/L$ approaches
  its $T=0$ limit exponentially in $1/T$ (because of the existence of
  an energy gap). We have an inflection point at $T\!=\!
  T_\mathrm{inf}^{(L)}$ (obtained from a cubic spline interpolation of
  $\xi_L/L$), that we regard as a proxy for the crossover scale
  $T^*_L$~\cite{thomas:11}. At low $T$ (discontinuous lines) we use
  less samples, see appendix~\ref{sect:parameters}.  {\bf Inset:} Size
  evolution of the inflection points $T_\mathrm{inf}^{(L)}$ (red full
  squares), compared to $T_{\xi_L/L=0.5}^{(L)}$ (open green
  circles). Data for binary model. As expected~\cite{parisen:11}, the
  two temperature scales decouple for large $L$.  {\bf Bottom:}
  $\xi_L/L$ vs. $T$ for the Gaussian model does not show any
  crossover.}
\label{fig:xi}
\end{figure}
Let us start with $\xi_L$. The Gaussian model,
Fig.~\ref{fig:xi}--bottom, displays the expected divergence upon
approaching $T=0$. In fact, the temperature where $\xi_L/L=x$, denoted
$T_{(\xi_L/L)=x}^{(L)}$ hereafter, decreases for larger sizes
[Eq.~(\ref{eq:FSS-xi}) predicts $T_{(\xi_L/L)=x}^{(L)}\sim
  L^{-1/\nu}$, see below].  As for the binary model, see
Fig.~\ref{fig:xi}--top and inset, its $\xi_L/L$ curves reflect the
different behaviors above and below the temperature scale
$L^{-\theta_S}$~\cite{thomas:11}. Here we do not investigate further
the $T\!\approx\! 0$ region nor this crossover.

\begin{figure}[t!]
\centering
\includegraphics[height=\columnwidth, angle=270]{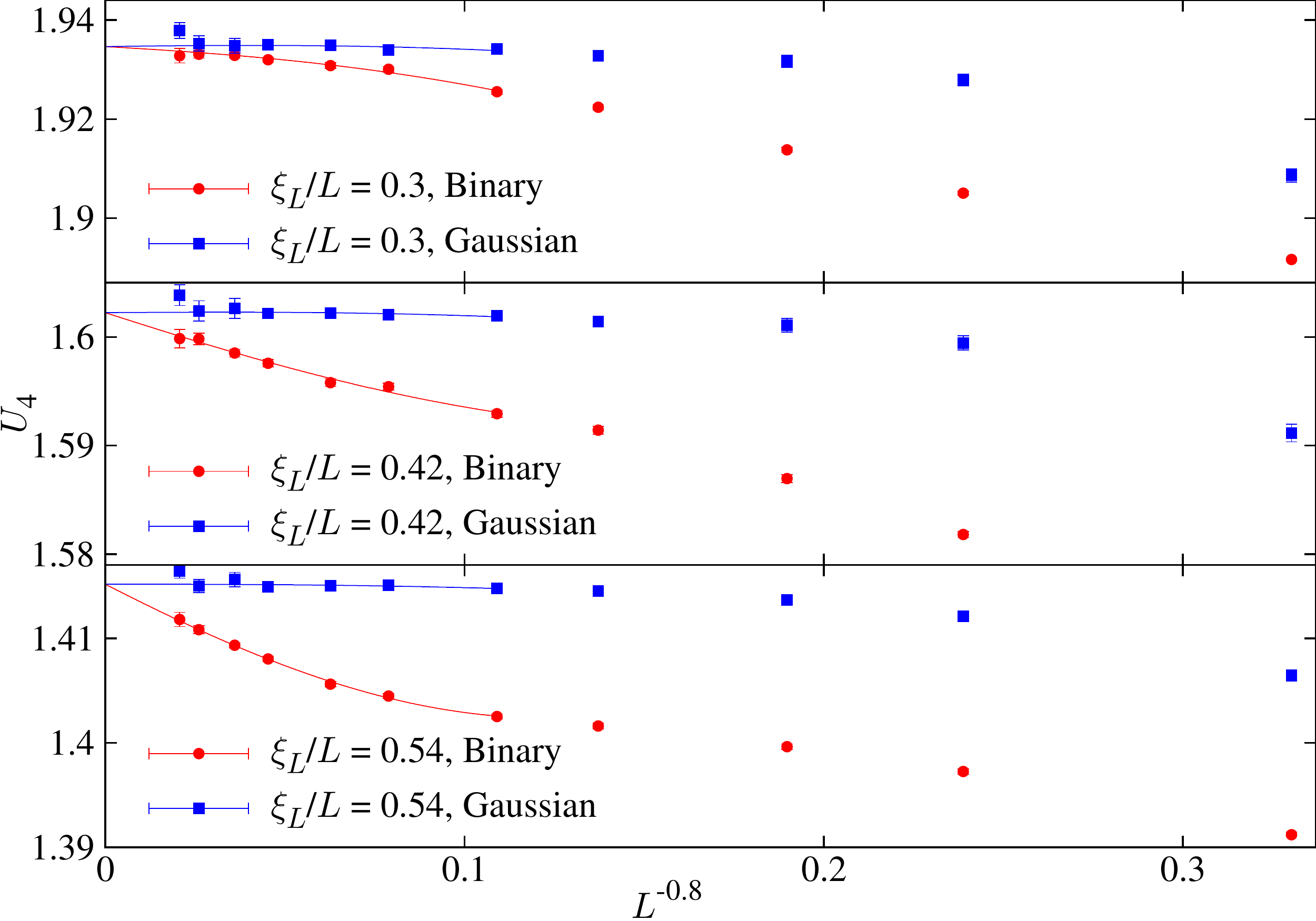}
\caption{(color online) Binder ratio $U_4$, Eq.~\eqref{eq:Binder}, at
  $T$ where $\xi/L=0.3$ ({\bf top}), 0.42 ({\bf center})
  and 0.54 ({\bf bottom}) as a function of $L^{-\omega}$ for the two
  models. The large $L$ limit is model independent. The $\omega$
  exponent and the solid lines were obtained from a joint fit to
  Eq.~\eqref{eq:omega-fit}.}
\label{fig:omega}
\end{figure}

Fortunately, universality emerges clearly if we bypass the temperature
dependency as done in Eqs.~(\ref{eq:FSS-xi},\ref{eq:FSS}).  $U_4$ at
$T_{\xi_L/L}^{(L)}$ reach an $\xi_L/L$-dependent universal limit for
large values of $L$, as shown in Fig.~\ref{fig:omega}. We
compute the corrections to scaling exponent $\omega$ from the behavior of $U_4$.
One expects corrections to the leading behavior:
\begin{equation}\label{eq:omega-fit}
U_4^{(L)}\big(T_{\xi_L/L}^{(L)}\big)=F_{U_4}({\textstyle\frac{\xi_L}{L}})+a({\textstyle\frac{\xi_L}{L}}) L^{-\omega}+
b({\textstyle\frac{\xi_L}{L}})L^{-(2-\eta)}\ldots\,.
\end{equation}
The amplitudes $a(\frac{\xi_L}{L}),b(\frac{\xi_L}{L})$
are model and $\xi_L/L$-dependent. 
If $\eta\!=\!0$
analytic corrections are ${\cal O}(L^{-2})$~\cite{amit:05}.

We fit together binary and Gaussian data to Eq.~\eqref{eq:omega-fit} by
standard $\chi^2$ minimization, imposing a common $F_{U_4}(\xi_L/L)$. The
goodness-of-fit estimator $\chi^2$ is computed with the full covariance
matrix, which limits the number of $\xi_L/L$-values that one may consider
simultaneously in the fit.

In our fit to Eq.~\eqref{eq:omega-fit} we include data for
$\xi_L/L=0.3, 0.42, 0.54$ and $L\geq L_\mathrm{min}$.  We impose two
requirements: (i) an acceptable $\chi^2/\mathrm{dof}$; (ii) stability
in the fitted parameters upon increasing $L_\mathrm{min}$. We obtain
$\omega=0.80(10)$ for $L_\mathrm{min}=16$, with
$\chi^2/\mathrm{dof}=23.9/26$. Interestingly, the amplitude
$a({\frac{\xi_L}{L}})$ for the Gaussian model is compatible with zero
for all values of $\xi_L/L$: the Gaussian model seems free of the leading
corrections to scaling~\footnote{Data for the Gaussian model can
  be fit as well with a sub-leading correction term $L^{-2\omega}$,
  rather than with the $L^{-2}$ term we use in
  Eq.~\eqref{eq:omega-fit}. With either sub-leading term we found that
  the leading corrections to the Gaussian data vanish within numerical
  accuracy.}.

As a control of systematic errors, we evaluated a second fit imposing
$b(\frac{\xi_L}{L})=0$ and, for the Gaussian data, also
$a(\frac{\xi_L}{L})=0$.  We obtained $\omega=0.69(5)$ for
$L_\mathrm{min}=32$ and $\chi^2/\mathrm{dof}=14.3/23$. Our
final estimate is
\begin{equation}
\omega=0.75(10)(5)\,.
\end{equation} 
(first is the  statistical error and second
the systematic one).

\section{The anomalous dimension.}
Previous investigations have never succeeded in computing
the anomalous dimension of the $2D$ spin glass. Our key idea is that
Eq.~\eqref{eq:FSS} implies $\eta=0$, provided that $\hat
u_h(T\!=\!0)\neq 0$ (traditional
methods cannot handle the prefactor $[\hat
  u_h(T)]^2$, see appendix~\ref{sect:traditional}).

\begin{figure}[t!]
\centering
\includegraphics[height=\columnwidth, angle=270]{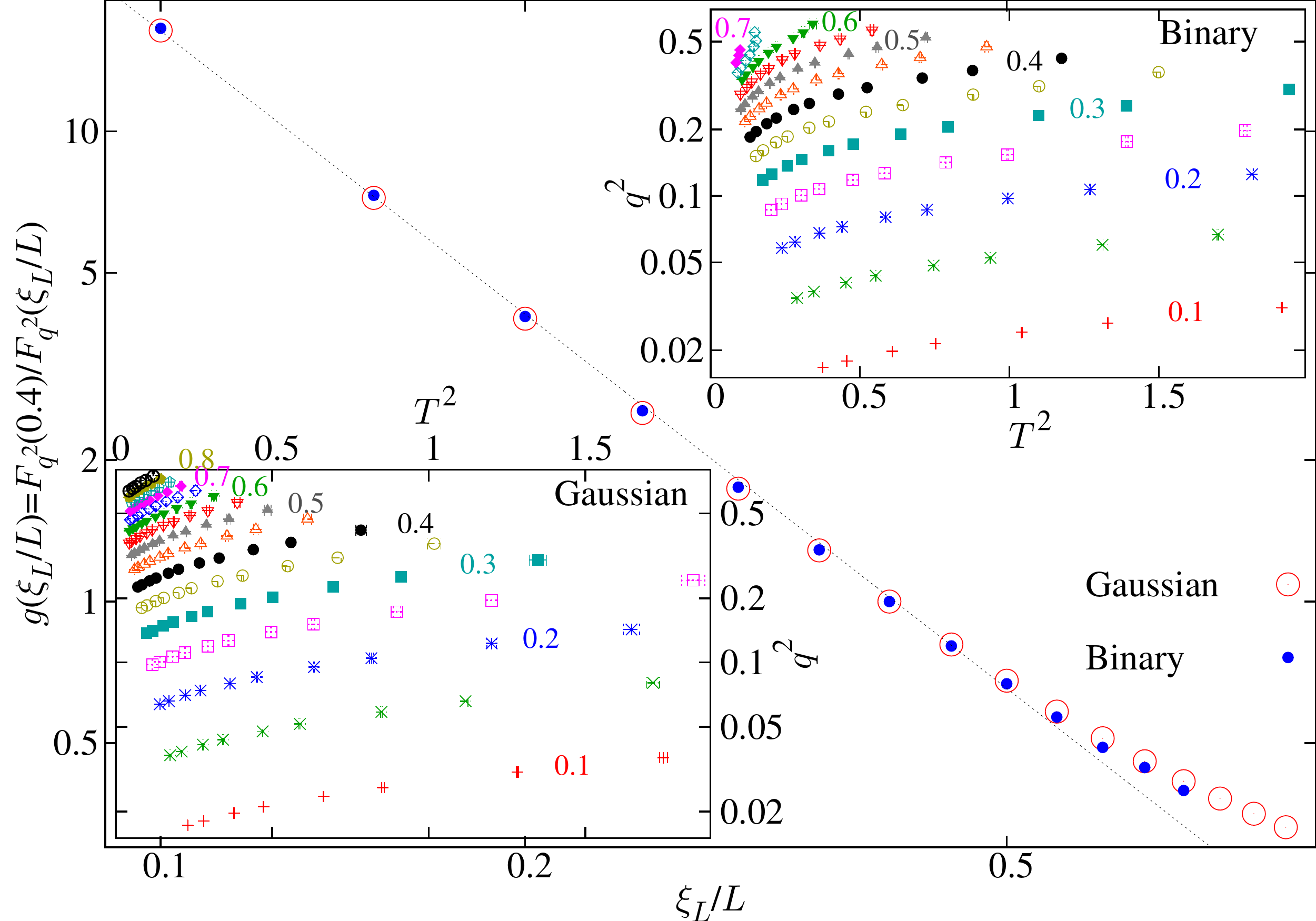}
\caption{(color online) Order parameter $\overline{\langle
    q^2\rangle}$ computed at fixed values of $\xi_L/L$ vs.
  $\big[T_{\xi_L/L}^{(L)}\big]^2$, for the binary ({\bf upper inset})
  and the Gaussian ({\bf lower inset}) models. {\bf Main:} universal
  scaling function $g(\xi_L/L)=F_{q^2}(0.4)/F_{q^2}(\xi_L/L)$,
  Eq.~\eqref{eq:FSS}, as computed for the Gaussian (empty symbols) and
  the binary (full symbols) models.  The function $g(x=\xi_L/L)$
  scales as $1/x^2$ for small $x$ (dashed line).}
\label{fig:scaling-gy}
\end{figure}

We focus on the temperature dependence of $\overline{\langle q^2\rangle}$, 
as computed at fixed $\xi_L/L$. For each $L$ we choose
$T=T_{\xi_L/L}^{(L)}$, see the two insets in
Fig.~\ref{fig:scaling-gy}. Eq.~\eqref{eq:FSS} tells that, apart
from a constant $F_{q^2}(\xi_L/L)$, the curves should be smooth
functions of $T^2$.

To compute the universal function $F_{q^2}(\xi_L/L)$ we
arbitrarily fix the scale $(\xi_L/L)=0.4$ (since, see
Fig.~\ref{fig:scaling-gy}, all our curves for
$\overline{\langle q^2\rangle}$ at fixed $\xi_L/L$ have some
temperature overlap with the curve for $(\xi_L/L)=0.4$). We fit 
to a quadratic polynomial in $T^2$ each curve $\overline{\langle
  q^2\rangle}$ at fixed $\xi_L/L$ for an interval $0< T^2 <
T^2_{\mathrm{max}, \xi_L/L}$, see appendix~\ref{sect:T-fits}. We compute
$g(\xi_L/L)\equiv F_{q^2}(0.4)/F_{q^2}(\xi_L/L)$ as the ratio of the two
$T^2$-fits, the one for a generic value of $\xi_L/L$ and the fit for
$(\xi_L/L)=0.4$, as evaluated at $T^2=T^2_{\mathrm{max}, \xi_L/L}/2$.

\begin{figure}[t!]
\centering
\includegraphics[height=\columnwidth, angle=270]{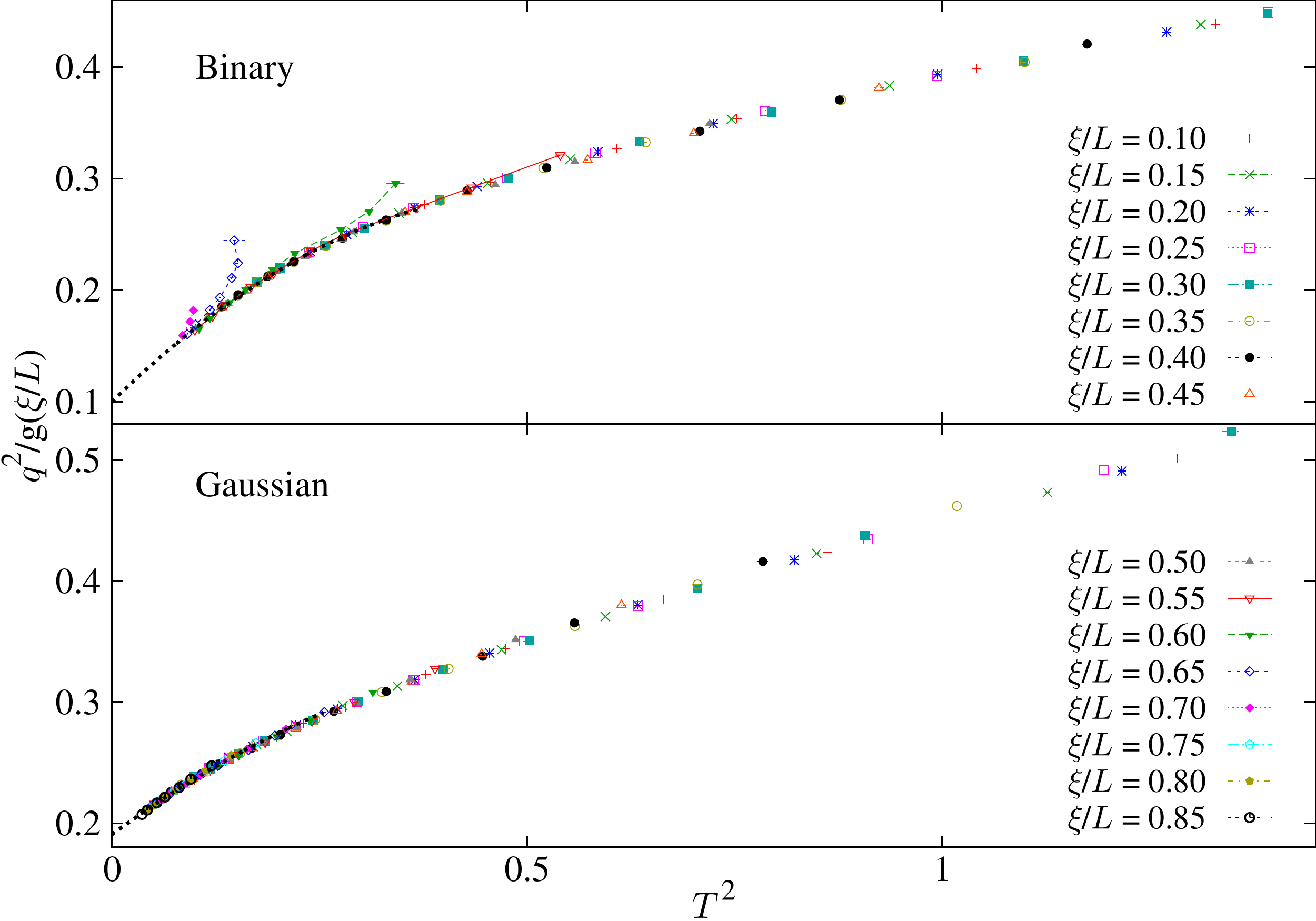}
\caption{(color online) Scaling field $[\hat u_h(T)]^2$ (from
  Eq.~\eqref{eq:FSS}) vs. $[T_{\xi_L/L}^{(L)}]^2$, as computed for the
  Gaussian ({\bf top}) and the binary ({\bf bottom}) models. The data
  collapses were obtained by multiplying the data in the two insets in
  Fig.~\ref{fig:scaling-gy} by the universal function $g(\xi_L/L)$,
  depicted in the main panel of Fig.~\ref{fig:scaling-gy}. The dots
  are for the extrapolation to $T^2=0$. The binary model data show
  the crossover between the $T=0$ (small $L$) and $T>0$ (large $L$)
  regimes (see Fig.~\ref{fig:xi} and
  Refs.~\cite{thomas:11,parisen:11}).}
\label{fig:q2-scaling}
\end{figure}

Our computation of the ratio $g(\xi_L/L)$ respects three
consistency tests: (i) $g(\xi_L/L)$ turns out to be
essentially model independent (Fig.~\ref{fig:scaling-gy});
(ii) $g(\xi_L/L)\sim (L/\xi_L)^{2}$ for small $\xi_L/L$
(Fig.~\ref{fig:scaling-gy}); (iii) the product of
$\overline{\langle q^2\rangle}$ at fixed $\xi_L/L$ with $g(\xi_L/L)$
produces $\xi_L/L$ independent curves.
(Fig.~\ref{fig:q2-scaling}).

Fig.~\ref{fig:q2-scaling} shows the (modified) scaling field $[\hat
  u_h(T)]^2 F_{q^2}(0.4)$. Given the $T^2$ fits  it is
straightforward to extrapolate $[\hat u_h(T)]^2 F_{q^2}(0.4)$ to $T^2=0$
(dashed lines in Fig.~\ref{fig:q2-scaling}). For both models
the extrapolation is non-vanishing (implying $\eta=0$). 

Finally, we obtain $\eta=0.00(2)$ from the scaling $g(x)\sim
x^{\eta-2}$ for small $x=\xi_L/L$ ($L\to\infty$ is taken at fixed $x$,
see appendix~\ref{sect:g-computation}).

\section{The thermal exponent.}\label{sect:nu}
The exponent $\nu$ has never been successfully computed for this
model~\footnote{The tentative estimate of ref.~\cite{rieger:96} was
  later found to be problematic \cite{rieger:97}}.
RG suggests that $1/\nu=-\theta$, where $\theta$ is
the stiffness exponent controlling the size scaling of the change in
the ground state energy when considering periodic and anti-periodic
boundary conditions. Accurate determinations of $\theta$ are available
for the Gaussian model: $-\theta=0.281(2)$~\cite{rieger:96},
$0.282(2)$~\cite{hartmann:01b}, $0.282(3)$~\cite{carter:02} and
$0.282(4)$~\cite{amoruso:03}. A computation for the
random anisotropy model yields $\theta=0.275(5)$~\cite{liers:07}. We
shall obtain results of comparable accuracy for $1/\nu$.  Due to the
strong cross-over effects suffered by the binary model (see
Fig.~\ref{fig:xi}) we estimate $1/\nu$ for the Gaussian
model only.

We base our analysis on the determination of
$T_{\xi_L/L}^{(L)}$. Even disregarding the
leading universal corrections to scaling (see above our
computation of
$\omega$), Eq.~(\ref{eq:FSS-xi}) predicts a rather complex
behavior, with
$\hat u_T(T_{\xi_L/L}^{(L)})= L^{-1/\nu}F_\xi^{-1}(\xi_L/L)$.
Inverting this relation, one obtains
$T_{\xi_L/L}^{(L)} = d_1^{(\xi_L/L)} L^{-1/\nu} + d_3^{(\xi_L/L)}
L^{-3/\nu}+ d_5^{(\xi_L/L)} L^{-5/\nu}+\ldots$. Since
$1/\nu\approx0.28$,
we expect annoying corrections to scaling due to the
non-linearity of the scaling fields. Were $\hat u_T(T)$ analytically
known, we could easily get rid of these corrections. We shall not achieve
this, but we shall get close to it.

\begin{figure}[t!]
\centering
\includegraphics[height=\columnwidth, angle=270]{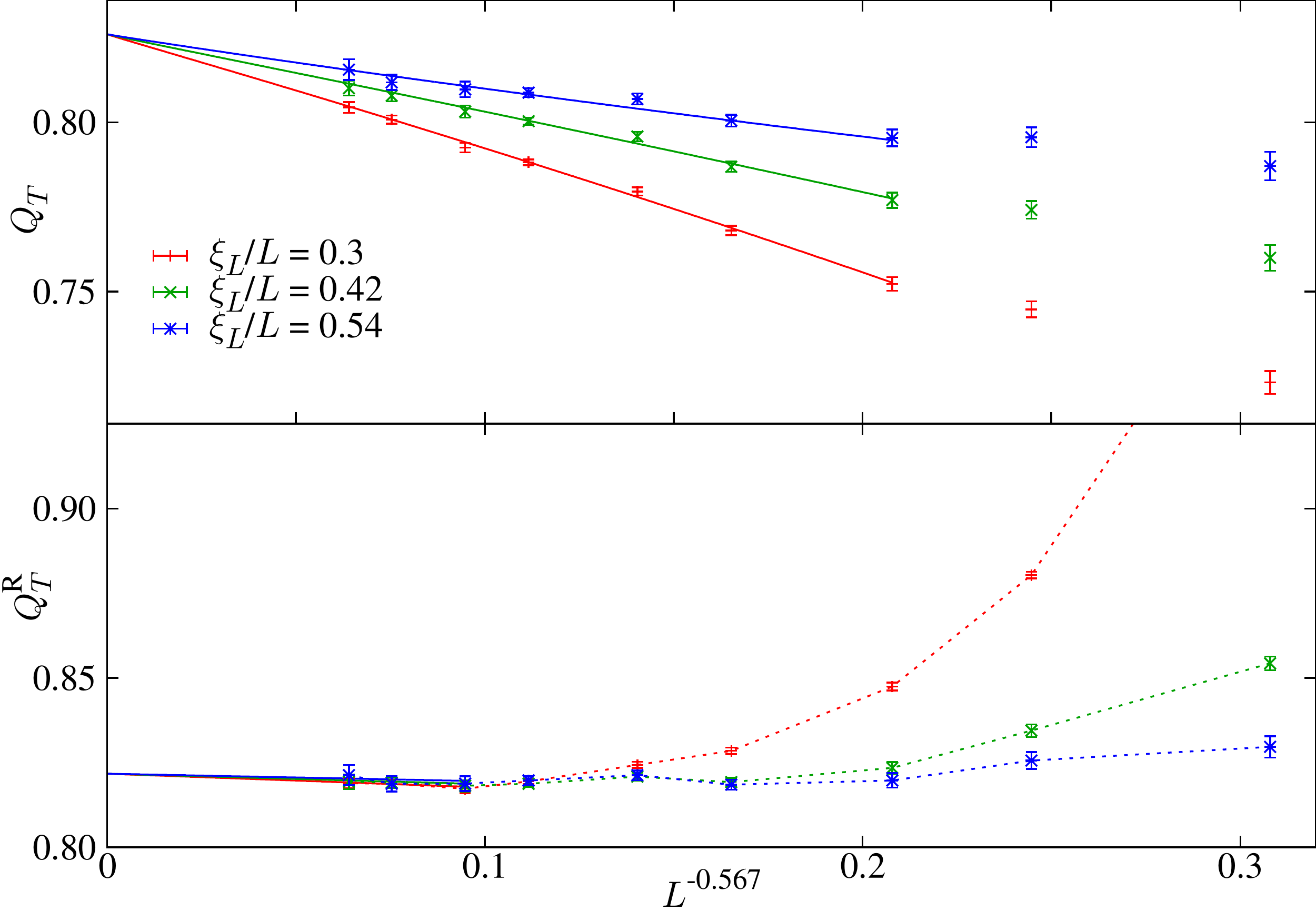}
\caption{(color online) Computing $\nu$ for the Gaussian model. Bare
  [{\bf top}, see Eq.~\eqref{eq:nu-fit-0}] and Renormalized [{\bf
      bottom}, Eq.~\eqref{eq:nu-fit-1} with $\hat u_3=-0.32$]
  temperature quotients at fixed $\xi_L/L (=0.3,0.42,0.54)$ as a
  function of $L^{-2/\nu}$. Continuous lines are our fits
  (see text), dotted lines are guides to eyes.}
\label{fig:nu_eff}
\end{figure}

In order to eliminate the unknown scaling function $F_\xi$, we compare couples
of lattices of size $L$ and $2L$:
\begin{equation}\label{eq:nu-fit-0}
Q_T(L)=\frac{T_{\xi_L/L}^{(2L)}}{T_{\xi_L/L}^{(L)}}=2^{-1/\nu}\,\frac{1+u_3 [T_{\xi_L/L}^{(L)}]^2+\ldots}{1+u_3 [T_{\xi_L/L}^{(2L)}]^2+\ldots}\,.
\end{equation}
In fact, see Fig.~\ref{fig:nu_eff}--top, scaling corrections are
strong, and strongly dependent on $\xi_L/L$.

We can alleviate the situation by introducing a renormalized
quotient
\begin{equation}\label{eq:nu-fit-1}
Q^{\mathrm{R}}_T(L)= \frac{T_{\xi_L/L}^{(2L)}}{T_{\xi_L/L}^{(L)}}\,
\frac{1+\hat u_3 [T_{\xi_L/L}^{(2L)}]^2}{1+\hat u_3 [T_{\xi_L/L}^{(L)}]^2}\,.
\end{equation}
Setting $\hat u_3=u_3$ we would have $Q^{\mathrm{R}}_T(L)=2^{1/\nu}+{\cal O}
(u_5 L^{-4/\nu})$. We have found that $\hat u_3=-0.32$ produces a
negligible slope: the remaining corrections in
Fig.~\ref{fig:nu_eff}--bottom are certainly of a different origin (either $u_5$
terms, analytic corrections to scaling, or even $L^{-\omega}$ terms).

We obtained a fit $Q^{\mathrm{R}}_T(L)=2^{1/\nu}+ d^{(\xi_L/L)} L^{-2/\nu}$
(i.e. we did \emph{not} assume $\hat u_3=u_3$) finding 
\begin{equation}
1/\nu=0.283(6)\,,\quad \chi^2/\mathrm{dof}=4.1/6\,\ (L_\mathrm{min}=64).
\end{equation}
Variations of $10\%$ in $\hat u_3$ change the $1/\nu$ estimate by one third of
the error bar. Furthermore, we can fit directly $Q_T(L)$, see
Fig.~\ref{fig:nu_eff}--top. In this case, we need to introduce corrections
quadratic in $L^{-2/\nu}$. We find a fair fit for $L_{\mathrm{min}}=16$ with
$1/\nu=0.275(9)$.

\section{Conclusions.} 
We have presented a high accuracy numerical simulation of the
Edwards-Anderson spin glass model in $2D$. We
consider systems with binary and Gaussian random couplings.
By focusing on renormalized quantities we are able to bypass the
peculiar temperature evolution dictated by the binary distribution. The
Binder ratios at fixed $\xi_L/L$ are fully compatible, in
the precision given by our small statistical errors, with a single
universality class. This analysis yields the first computation of the
leading corrections to scaling exponent $\omega$.  We identify the
non-linearity of scaling fields as the major obstacle that impeded so
far an accurate computation of critical quantities. We are able
to give strong numerical evidence that the anomalous dimension $\eta$
vanishes. We consider the temperature evolution for the
Gaussian distribution, which is free of cross-over effects. We obtain
a reliable direct estimate of $\nu$. Therefore, we are able to
provide a stringent test of the generally assumed equivalence
$\theta=-1/\nu$.

\section{Acknowledgments}

This work was partially supported by the Ministerio de Ciencia y
Tecnolog\'{\i}a (Spain) through Grant Nos. FIS2012-35719-C02, FIS2013-42840-P,
by the Junta de Extremadura (Spain) through Grant No. GRU10158 (partially
founded by FEDER).

\appendix
\section{Parameters of simulations and fits}\label{sect:parameters}
\subsection{Numerical simulations}
\begin{table}
\begin{ruledtabular}
\begin{tabular}{rrrrrr}
$L$\phantom{$^*$} & $N_\mathrm{samples}$ & $N_\mathrm{MCS}$& $N_\mathrm{T}$  & $T_\mathrm{min}$ & $T_\mathrm{max}$\\\hline\hline
4\phantom{$^*$} &  25\,600 & 320\,000 & 14& 0.20& 1.5\\
$4^*$ &  204\,800 & 80\,000 & 20& 0.72& 1.5\\
6\phantom{$^*$} &  25\,600 & 320\,000 & 14& 0.20& 1.5\\
$6^*$ &  204\,800 & 80\,000 & 20& 0.65& 1.5\\
8\phantom{$^*$} &  25\,600 & 320\,000 & 14& 0.20& 1.5\\
$8^*$ &  204\,800 & 80\,000 & 22& 0.60& 1.5\\
12\phantom{$^*$} &  25\,600 & 320\,000 & 14& 0.20& 1.5\\
$12^*$ &  204\,800 & 80\,000 & 19& 0.53& 1.5\\
16\phantom{$^*$} &  25\,600 & 320\,000 & 14& 0.20& 1.5\\
$16^*$ &  204\,800 & 80\,000 & 18& 0.47& 1.5\\
24\phantom{$^*$} &  25\,600 & 320\,000 & 14& 0.20& 1.5\\
$24^*$ &  204\,800 & 80\,000 & 16& 0.45& 1.5\\
32\phantom{$^*$} &  25\,600 & 128\,0000 & 14& 0.20& 1.5\\
$32^*$ &  204\,800 & 80\,000 & 18& 0.40& 1.5\\
48\phantom{$^*$} &  25\,600 & 1\,920\,000 & 27& 0.20& 1.5\\
$48^*$ &  204\,800 & 160\,000 & 27& 0.35& 1.5\\
64\phantom{$^*$} &  25\,600 & 640\,000 & 26& 0.25& 1.5\\
$64^*$ &  204\,800 & 240\,000 & 26& 0.35& 1.5\\
96\phantom{$^*$} &  102\,400 & 320\,000 & 49& 0.30& 1.5\\
128\phantom{$^*$} &  25\,600 & 640\,000 & 49& 0.30& 1.5\\
\end{tabular}
\end{ruledtabular}
\caption{Details of the numerical simulations for the binary model.
  We show the simulation parameters for each lattice size $L$.
  $N_\mathrm{samples}$ is the number of simulated samples (in bunches
  of 128 samples, due to multi-spin coding).  $N_\mathrm{T}$ is the
  number of temperatures that were used in parallel tempering, with
  maximum and minimum temperatures $T_{\mathrm{max}}$ and
  $T_{\mathrm{min}}$, respectively. In general, temperatures were
  evenly spaced. However some system sizes appear twice in the
  table. In fact, we performed some higher accuracy simulations,
  marked by a $^*$, aiming to increase the accuracy in the computation
  of $T_{\xi_L/L}^{(L)}$, the temperature where $\xi_L/L$ reaches a
  given prescribed value (see Fig. 1) and to improve the computation
  of $\omega$ (see Fig. 2). For those extended runs, we increased the
  number of temperatures in the region where $\xi_L/L >0.3$, in order
  to reduce the error for temperature interpolations. Finally,
  $N_\mathrm{MCS}$ is the number of Monte Carlo steps (MCS) used in
  each numerical simulation. Each MCS consisted of 10 Metropolis
  sweeps at fixed temperature, followed by a cluster
  update~\cite{houdayer:01} and by a Parallel Tempering step
  \cite{hukushima:96,marinari:98b}.} \label{tab:simulations_binary}
\end{table}

\begin{table}
\begin{ruledtabular}
\begin{tabular}{rrrrrr}
$L$ & $N_\mathrm{samples}$ & $N_\mathrm{MCS}$& $N_\mathrm{T}$  & $T_\mathrm{min}$ & $T_\mathrm{max}$\\\hline\hline
  4 & 204\,800 & 160\,000& 31 &0.1 &1.5\\
  6 & 204\,800& 160\,000& 31&0.1 &1.5\\  
  8 & 204\,800 & 160\,000& 31&0.1 &1.5\\  
 12 & 204\,800& 160\,000& 31&0.1 &1.5\\  
 16 & 204\,800& 160\,000 &31 &0.1 &1.5\\  
 24 & 204\,800 & 160\,000 &31 &0.1 &1.5\\  
 32 & 204\,800& 320\,000&31 &0.1 &1.5\\  
 48 & 204\,800& 160\,000&27 & 0.2 &1.5\\  
 64 & 25\,600& 320\,000&53 &0.2 &1.5\\
 96 & 25\,600& 480\,000 &41 &0.2 &0.7\\
 128 & 25\,600& 800\,000&41 &0.2 &0.7\\
\end{tabular}
\end{ruledtabular}
\caption{Simulation details for the Gaussian model,
  as in Table~\ref{tab:simulations_binary}. Here
  the number of samples $N_\mathrm{samples}$ is
  given by the number of random choices of the absolute values of the
  couplings times
  128 independent random choices
  of the coupling signs for each set of absolute values (see
  Sect.~\ref{sect:MSC-GAUSS}).}\label{tab:simulations_gaussian}
\end{table}

The parameters describing our multi-spin coding simulations are given
in Tables~\ref{tab:simulations_binary} and
~\ref{tab:simulations_gaussian}. We treat temperature as a continuous
variable, even if our data are obtained only in the temperature grid
where our Parallel Tempering simulations take place. We solved this
problem by using a standard cubic-spline interpolation. Note that data
for neighboring temperatures are statistically correlated (because we
use Parallel Tempering) which makes interpolation particularly easy in
our case.

\subsection{Temperature fits}\label{sect:T-fits}
The computation of the scaling field $\hat u_h(T)$ and of the scaling
function $F_{q^2}$, depicted in Figs. 3 and 4, is based on a
temperature fit.  For each prescribed value of $\xi_L/L$ and each
system size $L$, we considered $\overline{ \langle
  q^2\rangle}_{\xi_L/L}$ (namely the squared spin overlap as computed
at $T=T_{\xi_L/L}^{(L)}$, the temperature needed to have $\xi_L/L$
equal to its prescribed value in a system of size $L$). For each fixed
value of $\xi_L/L$ we fitted $\overline{\langle
  q^2\rangle}_{\xi_L/L}$, as computed for all our system sizes, to a
second order polynomial in $[T_{\xi_L/L}^{(L)}]^2$. The fits were
performed in the range $0<T^2 < T^2_{\mathrm{max},\xi_L/L}$. The
values of $T^2_{\mathrm{max},\xi_L/L}$ were obtained with a simple
algorithm: 1) For $\xi_L/L=0.1$ we took
$T^2_{\mathrm{max},\xi_L/L}=0.8$. 2) We increased $\xi_L/L$ in steps
of $0.05$. 3) At each such step, $T^2_{\mathrm{max},\xi_L/L}$ was
divided by $1.1$.

The above procedure has general validity.
However for the binary case at large
$\xi_L/L\geq 0.6$ our data are strongly affected by the crossover from
the $T>0$ to the  $T=0$ behavior~\cite{thomas:11,parisen:11}, illustrated in
Figs. 1 and 3. In order to avoid as much as
possible the effects of this crossover in the temperature window used
in the fit, we employed $T^2_{\mathrm{max},\xi_L/L}=0.19, 0.13$ and
$0.11$ for $\xi_L/L=0.6, 0.65$ and $0.7$, respectively. Also for these
three cases, the comparison with $\xi_L/L=0.4$ (needed to compute the
scaling function $g$ in Fig. 3) was done at $0.8
T^2_{\mathrm{max},\xi_L/L}$.

\section{Multi spin coding the Gaussian model}\label{sect:MSC-GAUSS}

This section is divided in two parts. We first explain how we
define the multi spin coding algorithm with Gaussian couplings
in~\ref{sect:algorithm}. Next, we assess
in~\ref{sect:effective-number} the statistical effectiveness of our
algorithm.

\subsection{The algorithm}\label{sect:algorithm}

It has been known for a long time how to perform the Metropolis update
of a single spin using only Boolean operations (AND, XOR, etc.),
provided that couplings are binary $J_{\boldsymbol x\boldsymbol
  y}=\pm1$, see e.g.~\cite{newman:99}. Besides, modern CPU perform
synchronously independent Boolean operations for all the bits in a
computer word.

Multi-spin coding is the fruitful combination of the above two
observations: one codes, and simulates in parallel, as many different
samples as the number of bits a word contains. Modern CPUs enjoy
streaming extensions that allow to code in a word 128 (or even
more) spins pertaining to the same site but to different samples. The
most efficient version of our programs turns out to be the one with
128-bits words.

The situation changes, of course, when the couplings $J_{\boldsymbol
  x\boldsymbol y}$ are drawn from a continuous distribution, such as a
Gaussian. In fact, we are not aware of working multi-spin coding
strategies when the coupling distribution is continuous. We explain
now how we circumvented this problem~\footnote{Another general
  solution is to use a discrete approximation to the Gaussian
  distribution, such as the Gaussian-Hermite
  quadrature~\cite{abramowitz:72}. For instance, in
  Refs.~\cite{leuzzi:09,janus:14b} a Gaussian-distributed magnetic
  field was simulated in this way.}.

Before describing our algorithm let us spell the standard Metropolis
algorithm, phrased in a somewhat unusual (but fully orthodox) way.
Imagine we are working at inverse temperature $\beta=1/T$. When updating site ${\boldsymbol x}$
we attempt to flip the spin $\sigma_{\boldsymbol x}\rightarrow -\sigma_{\boldsymbol x}$. Specifically,
\begin{enumerate}
\item
  We extract a random number  $R$  uniformly distributed in $[0,1)$.
\item
  We compute the energy change $\Delta E$ that the system would suffer
  if the spin $\sigma_x$ was flipped.  In our case, $\Delta E= 2
  \sum_{{\boldsymbol y} \text{ neighbor of } {\boldsymbol x}}
  J_{\boldsymbol x\boldsymbol y} \sigma_{\boldsymbol
    x}\sigma_{\boldsymbol y}$
\item
  We reject the spin flip only if
  $\exp(-\beta \Delta E) < R$. Otherwise, we flip the spin.
\end{enumerate}
So, we shall first get the random number $R$, then check if the actual
$\Delta E$ forces us to reject the spin-flip. Let us see how it works.

Let us call $N_{\boldsymbol x}$ the set of the four nearest neighbors
of ${\boldsymbol x}$ in the square lattice endowed with periodic
boundary conditions. For later use, let us also split the couplings
into their absolute values and their signs $J_{\boldsymbol
  x\boldsymbol y}= |J_{\boldsymbol x\boldsymbol y}|\;
\mathrm{sgn}(J_{\boldsymbol x\boldsymbol y})$. The crucial observation
is that for fixed $ |J_{\boldsymbol x\boldsymbol y}|$ the sum
\begin{equation}\label{eq:Sx}
  S_{\boldsymbol x} = \sum_{\boldsymbol y\in N_{\boldsymbol x}}
  \left|J_{\boldsymbol x\boldsymbol y}\right|\;
  \mathrm{sgn}\left(J_{\boldsymbol  x\boldsymbol y}\right)\;
  \sigma_{\boldsymbol x}\sigma_{\boldsymbol y}\,, 
\end{equation}
can only take $2^4=16$ different values, because each term of the sum
in Eq.~\eqref{eq:Sx} is a binary variable
[$\mathrm{sgn}(J_{\boldsymbol x\boldsymbol y}) \sigma_{\boldsymbol
    x}\sigma_{\boldsymbol y}=\pm 1$] and there are 4 neighboring sites
${\boldsymbol y}$. Of course, $S_{\boldsymbol x}=\Delta E/2$ (recall
the above description of the Metropolis algorithm).  Now, let us name
the $16$ possible values of $S_{\boldsymbol x}$ as
\begin{equation}
s_0<s_1<\ldots<s_7<0<s_8<s_9<\ldots<s_{15}\,.
\end{equation}
In fact, the symmetry of the problem ensures that $s_7=-s_8$,
$s_6=-s_9$, etc. Note also that having $s_i=0$ for some $i$, or
$s_i=s_k$ for a pair $i$ and $k$, are zero-measure events.

Let us chose an (arbitrary) ordering for the four neighbors: South,
East, North and West. We have $S_{\boldsymbol x}=s_{15}$ when the four
signs are $\{ \mathrm{sgn}(J_{\boldsymbol x\boldsymbol y})
\sigma_{\boldsymbol x}\sigma_{\boldsymbol
  y}\}_{15}=\{+1,+1,+1,+1\}$. Next, let us consider $s_{14}$. If the
weakest link (i.e. smallest $|J_{\boldsymbol x\boldsymbol y})|$)
corresponded to (say) the East neighbor, then the array yielding
$s_{14}$ would be $\{ \mathrm{sgn}(J_{\boldsymbol x\boldsymbol
  y})\sigma_{\boldsymbol x}\sigma_{\boldsymbol
  y}\}_{14}=\{+1,-1,+1,+1\}$.  The groups of four signs are ordered in
such away to produce decreasing values of the 16 $s_i$'s. The eight
groups $\{ \mathrm{sgn}(J_{\boldsymbol x\boldsymbol
  y})\sigma_{\boldsymbol x}\sigma_{\boldsymbol y}\}_{15},\ldots,\{
\mathrm{sgn}(J_{\boldsymbol x\boldsymbol y})\sigma_{\boldsymbol
  x}\sigma_{\boldsymbol y}\}_{8}$ deserve special attention: if the
current configuration takes one of these values, then the energy will
\emph{increase} upon flipping $\sigma_{\boldsymbol x}$. If the energy
increases we shall be forced to reject the spin-flip (unless the
random number turns out to be small enough).

With these definitions, the algorithm is easy to explain. 
We draw a random number $0\leq R<1$ with uniform
probability. The Metropolis update of site ${\boldsymbol x}$ at
inverse temperature $\beta=1/T$ can be cast as follows:
\begin{enumerate}
\item If $R < \mathrm{e}^{-2\beta s_{15}}$
  we flip the spin $\sigma_{\boldsymbol x}\rightarrow -\sigma_{\boldsymbol x}$.
\item If $\mathrm{e}^{-2\beta s_{15}} < R < \mathrm{e}^{-2\beta
  s_{14}}$ and the current
  configuration of the four signs turns out to be identical to the
  \emph{forbidden} array $\{ \mathrm{sgn}(J_{\boldsymbol x\boldsymbol
    y}) \sigma_{\boldsymbol x}\sigma_{\boldsymbol y}\}_{15}$ we
  let $\sigma_{\boldsymbol x}$ unchanged. Otherwise, we reverse the
  spin.
\item If $\mathrm{e}^{-2\beta s_{14}} < R < \mathrm{e}^{-2\beta
  s_{13}}$ we reverse $\sigma_{\boldsymbol x}$ unless the current
  configuration of the four signs is identical to one of the two
  configuration in the forbidden set: $\{ \mathrm{sgn}(J_{\boldsymbol
    x\boldsymbol y}) \sigma_{\boldsymbol x}\sigma_{\boldsymbol
    y}\}_{15}$ or $\{ \mathrm{sgn}(J_{\boldsymbol x\boldsymbol y})
  \sigma_{\boldsymbol x}\sigma_{\boldsymbol y}\}_{14}$.
\item If $\mathrm{e}^{-2\beta s_{13}} < R < \mathrm{e}^{-2\beta
  s_{12}}$, the forbidden set contains $\{
  \mathrm{sgn}(J_{\boldsymbol x\boldsymbol y}) \sigma_{\boldsymbol
    x}\sigma_{\boldsymbol y}\}_{15}$, $\{ \mathrm{sgn}(J_{\boldsymbol
    x\boldsymbol y}) \sigma_{\boldsymbol x}\sigma_{\boldsymbol
    y}\}_{14}$ and $\{ \mathrm{sgn}(J_{\boldsymbol x\boldsymbol y})
  \sigma_{\boldsymbol x}\sigma_{\boldsymbol y}\}_{13}$. We reverse
  $\sigma_{\boldsymbol x}$ unless the current signs configuration is
  contained in the forbidden set.
\item The same scheme apply to the other intervals, up to
   $\mathrm{e}^{-2\beta
  s_{8}}< R$. In this  extremal case, the forbidden set contains all
the energy-increasing configurations of the four signs: $\{ \mathrm{sgn}(J_{\boldsymbol
  x\boldsymbol y})\sigma_{\boldsymbol x}\sigma_{\boldsymbol
  y}\}_{15},\ldots,\{ \mathrm{sgn}(J_{\boldsymbol x\boldsymbol
  y})\sigma_{\boldsymbol x}\sigma_{\boldsymbol y}\}_{8}$.
\end{enumerate}
We can bypass the use of floating point arithmetics by using
a look up table. For each of the $L^2$ sites
of the system we need to keep in our table the eight probability
thresholds
$$\mathrm{e}^{-2\beta s_{15}}< \mathrm{e}^{-2\beta s_{14}}< \ldots
<\mathrm{e}^{-2\beta s_{8}}\;,$$
and the corresponding eight  \emph{sometimes forbidden} 
four-signs configurations 
\begin{eqnarray*}
\{ \mathrm{sgn}(J_{\boldsymbol x\boldsymbol y})
\sigma_{\boldsymbol x}\sigma_{\boldsymbol y}\}_{15}\,,\ \{
\mathrm{sgn}(J_{\boldsymbol x\boldsymbol y}) \sigma_{\boldsymbol
  x}\sigma_{\boldsymbol y}\}_{14}\,,\ \ldots &&\\
\ldots\,,\ \{ \mathrm{sgn}(J_{\boldsymbol
  x\boldsymbol y}) \sigma_{\boldsymbol x}\sigma_{\boldsymbol
  y}\}_{8}\,.&&
\end{eqnarray*}
The look-up table is entirely determined by
the absolute values of the couplings $|J_{\boldsymbol x\boldsymbol y}|$.

At this point, our multi-spin coding solution is straightforward. We
chose to code 128 different samples in each computer word. We set
randomly and independently the \emph{sign} of each of the $128\times
2\times L^2$ couplings, $\mathrm{sgn}(J_{\boldsymbol x\boldsymbol
  y})=\pm 1$ with $50\%$ probability. However, we only extract
$2\times L^2$ independent absolute values $ |J_{\boldsymbol
  x\boldsymbol y}|$ from the Gaussian distribution. This
$|J_{\boldsymbol x\boldsymbol y}|$ is common to all the the 128 bits
in the computer word that codes the bond between lattice sites
${\boldsymbol x}$ and ${\boldsymbol y}$.

\subsection{The effective number of samples}\label{sect:effective-number}

As far as we know, our multi-spin coding scheme is new and it
has never been tested. Therefore, it is useful
to investigate its effectiveness. 

Let us consider a Monte Carlo simulation long enough to make thermal
errors negligible as compared to sample to sample
fluctuations~\footnote{This situation is not desirable~\cite{ballesteros:98},
  but it is almost automatically enforced by the standard
  thermalization tests for spin-glasses~\cite{fernandez:08b}}.
Let us now simulate $N_S$ \emph{independent} samples, in
order to compute the expectation value $\overline{\langle O\rangle}$
for an observable $O$. For instance, $O$ could be the energy density
$e=H/L^2$, or the squared spin overlap $q^2$.

Our estimate will suffer from a statistical error $\Delta_O$ of
typical (squared) size
\begin{equation}\label{eq:error_independent}
\Delta_O^2 = \frac{\mathrm{Var}(O)}{N_S}\,,
\end{equation}
where $\mathrm{Var}(O)= \overline{\langle O\rangle^2}-\overline{\langle
  O\rangle}^2$ is the variance of $O$. 

We want to analyze a situation in which the coupling absolute values
$|J_{\boldsymbol x\boldsymbol y}|$ are fixed while we average over
many different coupling signs. It will be useful to recall some simple
notions about conditional probabilities (the same ideas were heavily
used in Refs.~\cite{janus:10,janus:14c}). Let $\langle
O\rangle_{|J|,\mathrm{sgn}(J)}$ be the thermal expectation of $O$ for a given
sample. We split the
couplings in their absolute values and
their signs $J_{\boldsymbol x\boldsymbol
  y}= |J_{\boldsymbol x\boldsymbol y}| \; \mathrm{sgn}(J_{\boldsymbol x\boldsymbol
  y})$.  The conditional expectation value of $\langle
O\rangle_{|J|,\mathrm{sgn}(J)}$, given the absolute values for the couplings, is 
\begin{equation}
E(\langle O\rangle |\; |J| ) = \frac{1}{2^{N_{\mathrm{B}}}} \sum_{\{\mathrm{sgn}(J)\}}\langle O\rangle_{|J|,\mathrm{sgn}(J)}\,,
\end{equation}
where $N_{\mathrm{B}}=2 L^2$ is the number of  bonds in the square lattice and
the sum extends to the $2^{N_{\mathrm{B}}}$ equally probable sign-assignments
for the couplings.  The
relationship with the standard expectation values is straightforward
\begin{equation}
E(O)\equiv\overline{\langle O \rangle}= \int D|J|\, E(\langle O\rangle |\; |J| )\,, 
\end{equation}
where $\int D|J|$ indicates the average taken with respect to the absolute value of the couplings.

The variance can be treated in a similar way. The variance induced by the absolute values is
\begin{equation}
\mathrm{Var}_{|J|}(O)=\int D|J|\, \Big(E(\langle O\rangle |\;|J|)\, -\, E(O)\Big)^2\,.
\end{equation}
Instead, the $|J|$-averaged variance induced by the signs is
\begin{eqnarray}
&&\mathrm{Var}_{\mathrm{sgn}(J)}(O)=\\[1mm]
&&\int D|J|\, \frac{1}{2^{N_{\mathrm{B}}}} \sum_{\{\mathrm{sgn}(J)\}} \, \Big(\langle O\rangle_{|J|,\mathrm{sgn}(J)} \,-\,E(\langle O\rangle |\;|J|)\Big)^2\,.\nonumber
\end{eqnarray}
It is straightforward to show that
\begin{equation}\label{eq:regla_de_suma}
\mathrm{Var}(O)=\mathrm{Var}_{|J|}(O)+\mathrm{Var}_{\mathrm{sgn}(J)}(O)\,.
\end{equation}
We are finally ready to discuss our multi-spin coding simulation. Imagine we simulate $N_{|J|}$ choices of the absolute values for the couplings.
Our squared statistical error is
\begin{equation}
\Delta_{O,\mathrm{MSC}}^2 = \frac{1}{N_{|J|}}\Big[\mathrm{Var}_{|J|}(O)+\frac{\mathrm{Var}_{\mathrm{sgn}(J)}(O)}{128}\Big]\,.
\end{equation}
However, the comparison with Eq.~\eqref{eq:error_independent} suggests us to define the effective number of samples in our 128 bits, $N_{\mathrm{eff},O}$, through
\begin{equation}\label{eq:error_MSC}
\Delta_{O,\mathrm{MSC}}^2=\frac{\mathrm{Var}(O)}{N_{|J|}\, N_{\mathrm{eff},O}} 
\end{equation}
The combination of Eqs.~\eqref{eq:regla_de_suma} and~\eqref{eq:error_MSC} tells us that
\begin{equation}
N_{\mathrm{eff},O}=128\,\frac{1+z}{128+z}\quad\text{where}\quad z=\frac{\mathrm{Var}_{\mathrm{sgn}(J)}(O)}{\mathrm{Var}_{|J|}(O)}\,.
\end{equation}
Therefore, the effective number of samples in our 128 bits computer word is bounded as
\begin{equation}
1< N_{\mathrm{eff},O} < 128\,.
\end{equation}
If the variance ratio $z$ is small, then $N_{\mathrm{eff},O}\approx 1$
and we will gain nothing by multi-spin coding. On the other hand, if
the statistical fluctuations induced by the signs dominate, $z$ will
be large and we shall approach to the optimal efficiency
$N_{\mathrm{eff},O} = 128$.

The problem to assess the effectiveness of our approach beforehand is
that estimating the variances $\mathrm{Var}_{|J|}(O)$ or
$\mathrm{Var}_{\mathrm{sgn}(J)}(O)$ is not easy. However, we can do it
by running  two different
kinds of numerical simulations. On the
one hand we can perform simulations with $N_S$ independent couplings.
On the other hand, we use multi-spin coding in a simulation with
$N_{|J|}$ independent choices of the absolute values for the
couplings. Numerical estimates of the statistical errors,
$\tilde\Delta_O$ and $\tilde\Delta_{O,\mathrm{MSC}}$, can be obtained
in a standard way. Then, Eqs.~\eqref{eq:error_independent}
and~\eqref{eq:error_MSC} tell us that
\begin{equation}\label{eq:Neff_empirical}
  N_{\mathrm{eff},O}\approx
  \frac{ \tilde\Delta_O^2}{\tilde\Delta_{O,\mathrm{MSC}}^2}\frac{N_S}{N_{|J|}}\,.
\end{equation}

\begin{table}
\begin{ruledtabular}
\begin{tabular}{ccccccccc}
$L$ & $\xi_L$ & $T$ & $N_S$ & $N_{|J|}$ & $N_{\mathrm{eff},e}$& $N_{\mathrm{eff},q^2}$& $N_{\mathrm{eff},\xi_L}$& $N_{\mathrm{eff},U_4}$\\\hline\hline
8&3.031(9) & 0.7 & 200 & 200 & 1.1 & 8.8 & 11.3& 11.2\\\hline
64&4.599(12)& 0.7& 200 & 200 & 1.4 & 8.0 &  7.0&  8.1\\\hline
8 & 8.581(19)& 0.2& 200 & 200 & 0.9 & 34.2 & 42.4&58.6\\\hline
48&35.86(4)& 0.2& 200 & 1600 & 1.4 & 89.2 &106.4 &110.6\\
\end{tabular}
\end{ruledtabular}
\caption{Numerical estimation of the effective number of independent
  samples in a 128 bits computer word, from
  Eq.~\eqref{eq:Neff_empirical}. We give results obtained under
  different dynamical conditions for the following observables:
  internal energy $N_{\mathrm{eff},e}$, squared overlap
  $N_{\mathrm{eff},q^2}$, correlation length
  $N_{\mathrm{eff},\xi_L}$, and Binder ratio
  $N_{\mathrm{eff},U_4}$.  We somehow abuse notation when
  applying Eq.~\eqref{eq:Neff_empirical} to quantities such as the
  correlation length $\xi_L$ or the Binder ratio $U_4$, which are
  computed as non-linear functions of mean values of direct
  observables. The statistical error in the computation
  of $N_\mathrm{eff}$ is below $10\%$.}\label{Table:MSC}
\end{table}

Some numerical experiments, described in Table~\ref{Table:MSC},
convinced us that our multi-spin coding is extremely useful when
computing long-distance observables, particularly when the correlation
length is large $\xi_L\gg 1$ and the system size increases. On the
other hand, when computing short distance observables (such as the
internal energy), $N_{\mathrm{eff},O}$ turns out to be disappointingly
close to one. Fortunately, for long-distance quantities, such as the Binder parameter at $\xi\approx 36$, we have
an effective number of samples as large as
$N_{\mathrm{eff},U_4}\approx 111$.

\section{Computing the anomalous dimension}\label{sect:g-computation}

We have seen that
$$
\overline{\langle q^2\rangle} = [\hat u_h(T)]^2
F_{q^2}(\xi_L/L)\,,\ g(\xi_L/L)=\frac{F_{q^2}(0.4)}{F_{q^2}(\xi_L/L)}\,.
$$
Let us define $x\equiv\xi_L/L$. The universal scaling function
$g(x)$ was depicted in Fig. 3. We shall employ it
here, to obtain a quantitative bound on the anomalous dimension
$\eta$. 

If we take the $L\to\infty$ limit
at fixed $x$, for small $x$  we obtain
the scaling law
\begin{equation}
  \label{eq:g-scaling}
  g(x) \propto \frac{1}{x^{2-\eta}}\,.
\end{equation}

\begin{figure}[t!]
\centering \includegraphics[height=\columnwidth, angle=270]{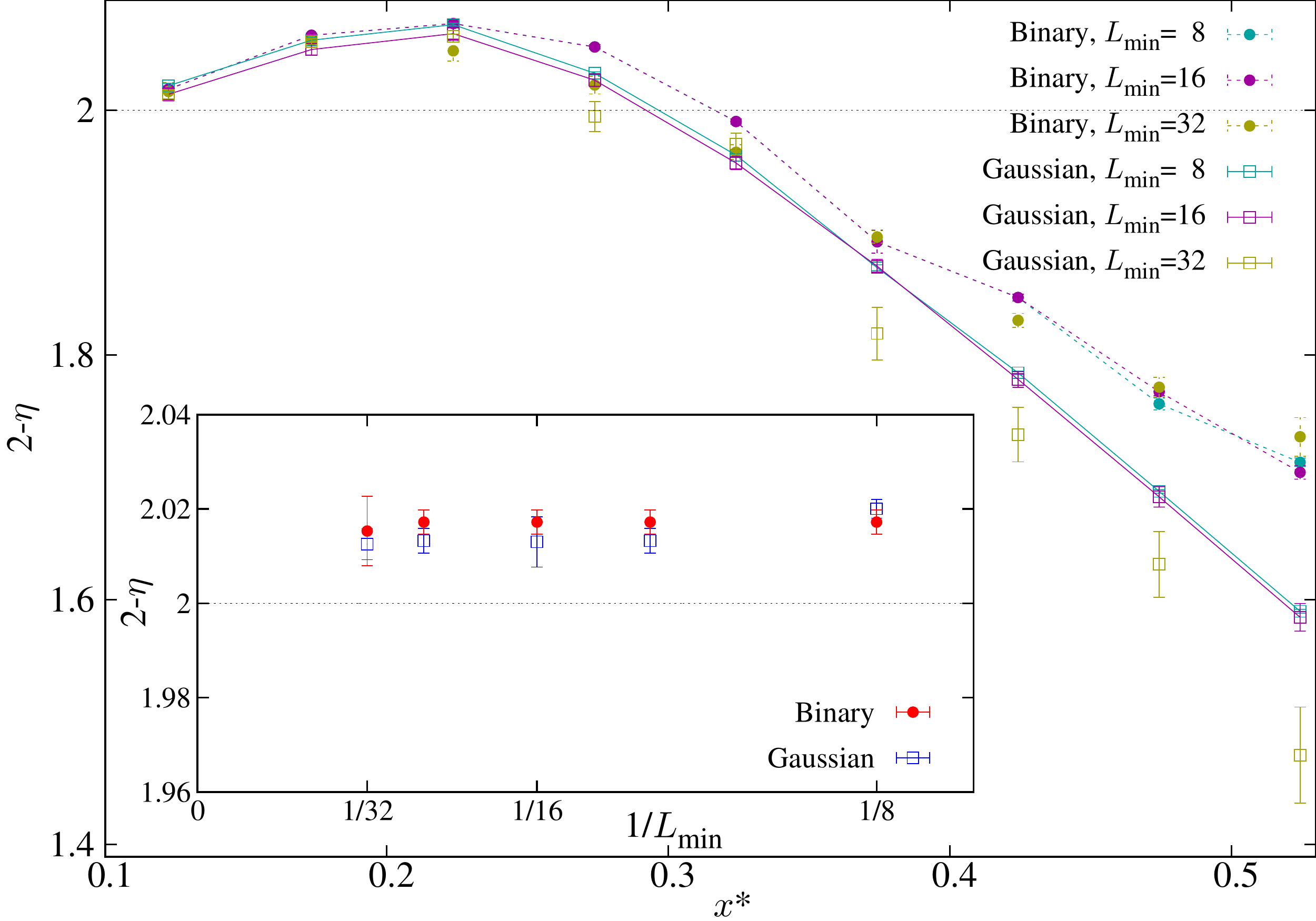}
\caption{Effective value of $2-\eta$ as obtained from
  Eq.~\eqref{eq:effective-eta} versus $x^*$ (which is the geometric
  mean of the two values of $\xi_L/L$ involved in the computation of
  $\eta$). We show estimations for several values of the minimal size
  included in the analysis, $L_\mathrm{min}$. Data for the binary
  model obtained with the same value of $L_\mathrm{min}$ are connected
  by dashed lines (continuous lines in the case of Gaussian
  distributed couplings). {\bf Inset:} For the smallest argument $x^*$
  that we reach in our simulations, we investigate the dependency of
  $2-\eta$ on $L_\mathrm{min}$.  }
\label{fig:effective-eta}
\end{figure}

Our procedure is as follows. We first determine $g(x,L_\mathrm{min})$
by computing the scaling function $g(x)$ as explained before, but
restricting the analysis to data from system sizes $L\geq
L_\mathrm{min}$. We then consider pairs of arguments $x_1$ and $x_2$
(consecutive points in the $x$ grid where we compute $g(x)$, see
Fig. 3) and obtain the effective estimators
\begin{equation}\label{eq:effective-eta}
2-\eta(x^*)= \frac{\log [g(x_1,L_\mathrm{min})/g(x_2,L_\mathrm{min})]}{\log [x_2/x_1]}\,,\ x^*\equiv\sqrt{x_1x_2}\,,
\end{equation}
that are shown in Fig.~\ref{fig:effective-eta}.

The estimations depicted in Fig.~\ref{fig:effective-eta} depend on
everything they could: on the disorder distribution, on
$L_\mathrm{min}$ and on $x^*$.  However, for small $x^*$ the
dependency on $L_\mathrm{min}$ and on the disorder distribution become
negligible within our better than $1\%$ accuracy (see
Fig.~\ref{fig:effective-eta}---inset)~\footnote{For $L_\mathrm{min}=
  32$, the automated selection of $T^2_{\mathrm{max},\xi_L/L}$ for the
  fits discussed in section~\ref{sect:T-fits} does not result into a
  good data collapse. For instance, for Gaussian couplings,
  $\xi_L/L=0.1$ and $L_\mathrm{min}\geq 32$ one needs to chose
  $T^2_{\mathrm{max},\xi_L/L}=0.53$ (rather than 0.8, as we choose for
  smaller $L_\mathrm{min}$)}.

It is obvious from Fig.~\ref{fig:effective-eta} that effects from
different origin compete: statistical errors and systematic errors due
to $x^*$ been too large (or to $L_\mathrm{min}$ being too
small). However, we have an additional hint: we expect $\eta=0$ for
the Gaussian model. But we see identical $1\%$ deviations from
$2-\eta=2$ for Gaussian and for binary couplings. Thus we regard the
small difference in the inset in Fig.~\ref{fig:effective-eta} as an
estimation of the combined errors (systematic and statistical) that we
suffer. We can safely summarize our findings as
\begin{equation}
|\eta_{\text{binary}}|<0.02\,.
\end{equation}


\section{Traditional analysis}\label{sect:traditional}

\begin{figure}[t!]
\centering \includegraphics[height=\columnwidth, angle=270]{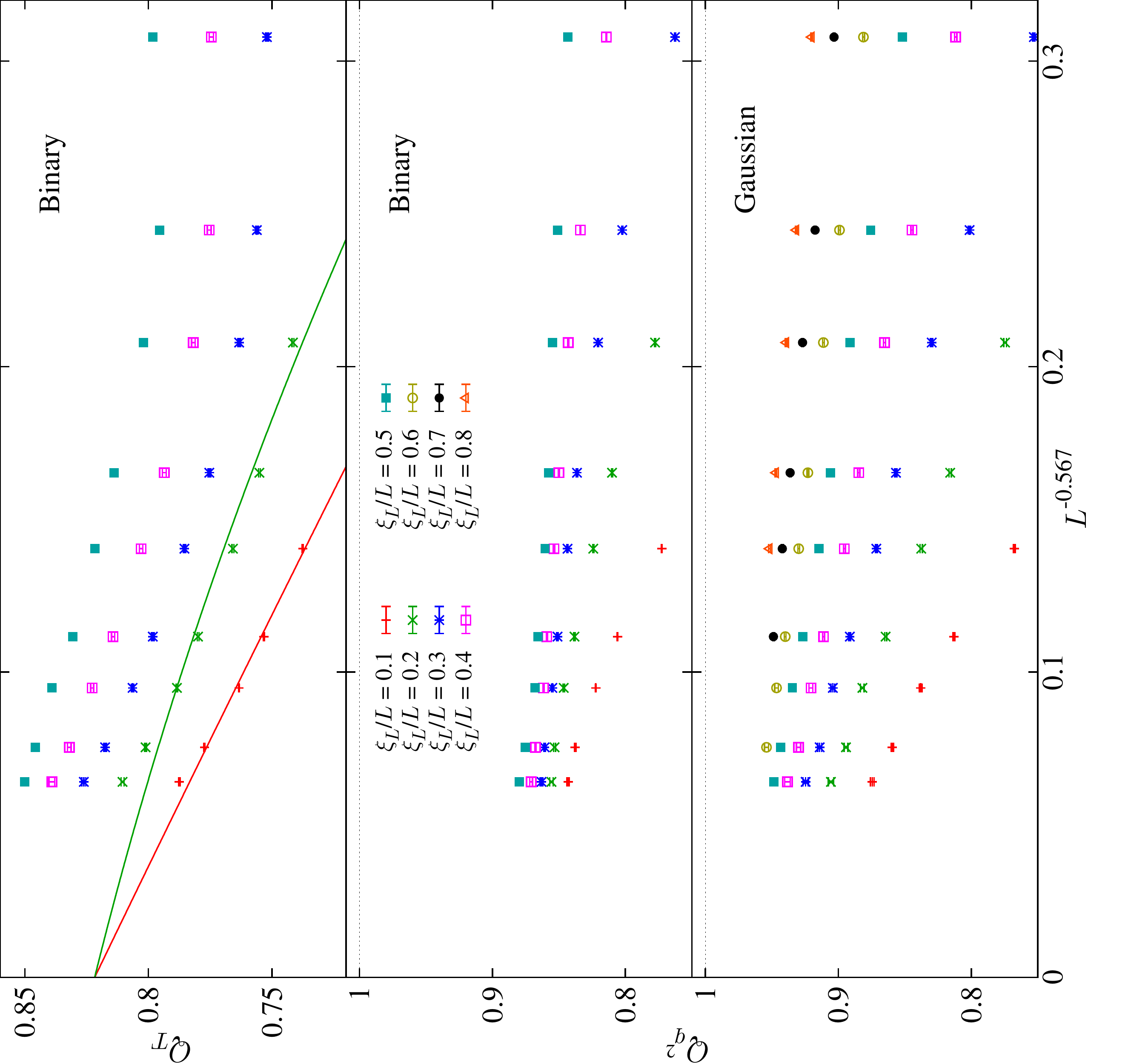}
\caption{The effective, size dependent
  critical exponents $\nu$ [{\bf Top:} binary model. $Q_T$ is defined
    in Eq.~\ref{eq:nu}.] and  the anomalous dimension $\eta$
    ({\bf Middle:} binary model. {\bf Bottom:} Gaussian model). The quotient
  $Q_{q^2}$ is defined in Eq.~\eqref{eq:eta} and analyzed in
  Eq.~\eqref{eq:eta-2}.}
\label{fig:mal}
\end{figure}

For sake of completeness, we include here the results of a traditional
analysis, based on scaling laws as a function of the system
temperature. These results give a flavor of
how severe are the problems caused by the non-linear scaling fields.

The difficulties encountered in the computation of the thermal
exponent $\nu$ are explained in Sect.~\ref{sect:nu}.
One can compute it from the comparison of temperatures
$T_{\xi_L/L}^{(L)}$ for lattices $L$ and $2L$:
\begin{equation}\label{eq:nu}
Q_T(L)=\frac{T_{\xi_L/L}^{(2L)}}{T_{\xi_L/L}^{(L)}}=2^{-1/\nu} (1+\ldots)\,.
\end{equation}
When computing this ratio for the Binary model, see
Fig.~\ref{fig:mal}--top, the scaling corrections come from a number
of different source.
We have, of course, the corrections due to the scaling field
$\hat u_T$ that were discussed in Sect.~\ref{sect:nu}. Yet, we also have
strong corrections of order ${\cal O}(L^{-\omega})$ [instead, for the
  Gaussian model we are fortunate to have tiny, probably negligible,
  ${\cal O}(L^{-\omega})$ corrections, see Fig. 2]. We also have to deal with the crossover between $T=0$ and
$T>0$ behaviors~\cite{thomas:11,parisen:11} (for a fixed variation
range of $L$, the crossover appears when increasing $\xi_L/L$). In
fact, we know that some of these scaling corrections are of similar
magnitude: those arising from $\hat u_T$ should be of order
$L^{-2/\nu}$ with $1/\nu=0.283(6)$ while
$\omega=0.75(10)(5)$. Disentangling the effects of the three sources
of corrections to scaling will require a strong analytical
guidance. Probably, simulating much larger systems, which is
possible using special methods~\cite{thomas:13}, will be useful.

As for the anomalous dimension, the traditional approach would start
from the quotients of $\overline{\langle q^2 \rangle}$ at fixed
$\xi_L/L$, as computed for $L$ and $2L$:
\begin{equation}\label{eq:eta}
Q_{q^2}(L)=\frac{\overline{\langle q^2 \rangle}(2L,T_{\xi_L/L}^{(2L)})}
{\overline{\langle q^2 \rangle}(L,T_{\xi_L/L}^{(L)})}\,.
\end{equation}
Barring scaling corrections, this quotient should behave as
$2^{-\eta}$. Therefore, for very large $L$, $Q_{q^2}(L)$ should
tend to one. The reason for this unfavorable behavior is that (ignoring all sort of
scaling corrections) this ratio actually behaves as
\begin{equation}\label{eq:eta-2}
  Q_{q^2}(L)=2^{-\eta}\Bigg(
  \frac{\hat u_h(T_{\xi_L/L}^{(2L)})}{u_h(T_{\xi_L/L}^{(L)})}
  \Bigg)^2\,.
\end{equation}
In fact, in the thermodynamic limit the two temperatures
$T_{\xi_L/L}^{(2L)}$ and $T_{\xi_L/L}^{(L)}$ tend to $T=0$, making the
ratio of scaling fields in Eq.~\eqref{eq:eta-2} irrelevant. However,
our data are far away from this limit, as shown in Fig. 4.

In fact, we know that $T_{\xi_L/L}^{(2L)}< T_{\xi_L/L}^{(L)}$ and that
$\hat u_h$ is an increasing function (recall again Fig. 4). It follows that the ratio of scaling functions in
Eq.~\eqref{eq:eta-2} is smaller than one, which mimics a slightly
positive \emph{effective} anomalous dimension, see
Fig~\ref{fig:mal}--middle and bottom.

\bibliographystyle{apsrev4-1}
\bibliography{../biblio.bib}

\end{document}